\documentclass[12pt,thmsa]{article}
\usepackage{amssymb}

\usepackage{sw20lart}

\input{tcilatex}
\begin{document}

\author{Emilio Santos \and Departamento de F\'{i}sica. Universidad de Cantabria.
Santander. Spain \and email: santose@unican.es}
\title{The cosmological constant problem or how the quantum vacuum drives the slow
accelerating expansion of the Universe }
\date{March, 13, 2018 }
\maketitle

\begin{abstract}
I argue that a solution to the cosmological constant problem is to assume
that the expectation value of the quantum vacuum stress-energy tensor is
proportional to the metric tensor with a negative energy density and
positive pressure. This assumption is confirmed by an explicit calculation
of the vacuum expectation for the free electromagnetic and Dirac fields of
quantum electrodynamics. As a consequence the metric of the universe might
correspond to a FLRW with accelerating expansion only after averaging over
large scales, but at small scales it gives rise to an extremely rapid
fluctuation between expansion and contraction in every small region, with
different phases in different points. The vacuum stress-energy tensor has
fluctuations that lead to short periods of expansion. A calculation with
plausible approximations leads to an estimate of the accelerating expansion
that fits in the observed value.
\end{abstract}

\section{ Introduction}

The Cosmological Constant Problem\cite{Weinberg} arises because, according
to standard wisdom, the quantum vacuum should give rise to an effective
cosmological term, in Einstein\'{}s equation of general relativity, about $%
10^{123}$ times too big (see Wang et al.\cite{Wang} and references therein).
After the discovery that the Universe is in accelerating expansion\cite
{expansion} the problem has turned out more dramatic. Indeed the
accelerating expansion is attributed to a socalled dark energy, but there is
increasing evidence that dark energy is equivalent to a cosmological
constant.

Actually the problem arises as a consequence of the following three
hypotheses (units $c=
\rlap{\protect\rule[1.1ex]{.325em}{.1ex}}h%
=1$ are used throughout):

H1- The stress-energy tensor of the quantum vacuum is proportional to the
metric tensor, therefore equivalent to a cosmological term. This means that
the vacuum pressure, $P^{vac},$ is isotropic and opposite of the vacuum
energy density, $\rho ^{vac},$ if both are measured in a local Minkowski
space. That is 
\begin{equation}
P^{vac}=-\rho ^{vac}.  \label{1}
\end{equation}
The reason for this hypothesis is the assumed Lorentz invariance of the
vacuum.

H2. The vacuum energy density is positive, that is 
\begin{equation}
\rho ^{vac}>0.  \label{2}
\end{equation}

H3. The absolute value of the energy density is of order the Planck density,
that is 
\begin{equation}
\left| \rho ^{vac}\right| \sim \rho _{Planck}=\frac{1}{%
\rlap{\protect\rule[1.1ex]{.325em}{.1ex}}h%
G^{2}}\sim 10^{125}\rho _{B},  \label{3}
\end{equation}
where $\rho _{B}$ is the mean baryonic density of the universe. The reason
for this hypothesis is that a straightforward calculation of the energy of
the quantum vacuum fields gives a divergent result and there is no natural
cutoff except at the Planck scale.

If we ignore, for simplicity, the (baryonic and dark) matter content of the
universe then hypothesis $H1$ and $H2$ lead to a metric that may be written 
\begin{equation}
ds^{2}=-dt^{2}+a(t)(dx^{2}+dy^{2}+dz^{2}),  \label{4}
\end{equation}
where the parameter $a(t)$ fulfils 
\begin{equation}
\frac{d^{2}a}{dt^{2}}=-\frac{4\pi G}{3}\left( \rho ^{vac}+3P^{vac}\right) a=%
\frac{8\pi G}{3}\rho ^{vac}a,  \label{5}
\end{equation}
that implies 
\begin{equation}
a(t)=a(0)e^{Ht},H\equiv \pm \sqrt{\frac{8\pi G}{3}\rho ^{vac}},  \label{6}
\end{equation}
where the choice of an initial positive value for the (Hubble parameter) $H$
leads to an accelerating expansion of the universe. Finally if we accept
hypothesis $H3$ the acceleration is huge, many orders greater than observed.
The conclusion changes but slightly if we take matter (baryonic and dark)
into account. This is the cosmological constant problem.

It is obvious that in order to escape from the problem one at least of the
three above hypotheses should be wrong. For some people the solution is to
reject $H3$ by the simple expedience of assuming that the huge vacuum energy
density is an artifact of the quantization, but the vacuum does not
contribute any energy or pressure. Indeed this was implicitly assumed by
Einstein when he introduced in 1917 a cosmological constant in his
equations, in order to allow for a static universe, that he later rejected
when the Hubble expansion was discovered. However there are good reasons to
assume that the quantum vacuum is not devoid of energy, for instance
phenomena like the Lamb shift or the Casimir effect. Also if the
accelerating expansion of the universe is not a quantum vacuum effect, then
a cosmological constant should be introduced, say by hand, without any
fundamental reason. A weaker assumption than putting $\rho ^{vac}=0$ is to
include a cutoff at energies much smaller than Planck's, eq.$\left( \ref{3}%
\right) .$ A proposal was made by Zeldovich in 1976\cite{Zel} substituting
the following for eq.$\left( \ref{3}\right) $%
\begin{equation}
\rho ^{vac}=\frac{Gm^{6}}{
\rlap{\protect\rule[1.1ex]{.325em}{.1ex}}h%
^{4}}.  \label{7}
\end{equation}
This fits in the needed dark energy if $m$ is of order the pion mass, but no
good reason is known for that.

In a recent paper by Q. Wang, Z. Zhu and W. G. Unruh\cite{Wang} the authors
reject hypothesis $H1$. In the following I summarize the arguments of that
paper that are relevant for us. They recognize that for quantum vacuum the
expectation 
\begin{equation}
\left\langle P^{vac}+\rho ^{vac}\right\rangle =0  \label{8}
\end{equation}
is the standard lore but claim that the stress-energy tensor has
fluctuations which are of the same order as the expectation value, and thus
semiclassical gravity which uses the expectation value of the stress-energy
tensor as the source for the gravitational field is a bad approximation.
Thus also the idea that the metric is sufficiently smooth at short scales
that one can apply Lorentz symmetry arguments to the stress-energy tensor is
highly suspect. On much longer scales than the cutoff these fluctuations
average out and the effective spacetime becomes approximately Lorentz
invariant, that is Lorentz invariance is an emergent phenomenon. Then they
substitute for eq.$\left( \ref{1}\right) $ the following 
\begin{equation}
\rho ^{vac}>-3P^{vac},  \label{9}
\end{equation}
see their eq.(42). Taking hypothesis $H2$ into account eq.$\left( \ref{9}%
\right) $ leads to an approximate metric similar to eq.$\left( \ref{4}%
\right) ,$ but the parameter $a$ fulfils 
\begin{equation}
\frac{d^{2}a}{dt^{2}}=-\Omega ^{2}a,\Omega ^{2}=\frac{4\pi G}{3}\left( \rho
^{vac}+3P^{vac}\right) >0,  \label{10}
\end{equation}
where now the parameter $a$ should depend on space coordinates, that is $%
a=a\left( x,y,z,t\right) .$ Thus the quantum vacuum would not produce
expansion of the universe, but rapid oscillation between expansion and
contraction in every small region. On the latter part of the article\cite
{Wang} the authors claim to prove that in the said oscillations the
expansion dominates slightly over the contraction thus driving the slow
accelerating expansion of the universe. I shall not comment here on that
part of the paper.

I propose that the hypothesis to be rejected is $H2$. The positivity of
vacuum energy density is a cherised assumption but there is a fundamental
reason to reject it, namely that an actual calculation of the stress-energy
tensor of the quantum vacuum leads to eq.$\left( \ref{8}\right) $ with $%
\left\langle \rho ^{vac}\right\rangle <0,$ at least for the fields of
quantum electrodynamics, as is shown in the next section. In this case eq.$%
\left( \ref{10}\right) $ holds true, that is the quantum vacuum
stress-energy produces rapid fluctuations in the metric, rather than
accelerating expansion. I shall show that the fluctuations of the vacuum
stress-energy tensor give rise to an effective slow expansion in addition to
the metric fluctuations. Thus my conslusion is similar to the one of Wang et
al. \cite{Wang}, but from different assumptions.

In order to study the consequences of rejecting hypothesis $H2$ I will
consider a metric of FLRW type with spatial dependence, that is

\begin{equation}
ds^{2}=-dt^{2}+a^{2}(x,y,z,t)\left[ dx^{2}+dy^{2}+dz^{2}\right] .
\label{2.10}
\end{equation}
Hence finding the components of the Einstein equation is straightforward
(see e.g. \cite{Wang}), but only two are needed for our purposes here. One
of them is the following, that turns out to be independent of the space
coordinates (compare with eq.$\left( \ref{5}\right) $),

\begin{equation}
\frac{d^{2}a}{dt^{2}}\equiv \ddot{a}=-\frac{4\pi G}{3}\left( \rho +3P\right)
a,  \label{2.11}
\end{equation}
and it holds true for any diagonal stress tensor, not necessarily fulfilling
hypothesis $H1$. The second equation is the following 
\begin{equation}
4\pi G(\rho +P)=0=-2\frac{\ddot{a}}{a}+2\left( \frac{\dot{a}}{a}\right) ^{2}+%
\frac{1}{3a^{2}}\left( \frac{\nabla a}{a}\right) ^{2}-\frac{2}{3a^{2}}\left( 
\frac{\nabla ^{2}a}{a}\right) ,  \label{2.12}
\end{equation}
where I do assume hypothesis $H1$. The interesting and well known result is
that when $\rho +3P<0$ there is a solution of both eqs.$\left( \ref{2.11}%
\right) $ and $\left( \ref{2.12}\right) $ with the parameter $a$ independent
of space cordinates, that is given by eqs.$\left( \ref{4}\right) $ to $%
\left( \ref{6}\right) ,$ meaning an exponential homogeneous expansion of
space. It may be shown that this solution fulfils also the remaining
Einstein equations for the metric eq.$\left( \ref{2.10}\right) $. In sharp
contrast, if $\rho +3P>0$ the solution of eq.$\left( \ref{2.11}\right) $ is
of the form of eq.$\left( \ref{10}\right) ,$but no solution of both eqs.$%
\left( \ref{2.11}\right) $ and $\left( \ref{2.12}\right) $ exists with $a$
independent of space coordinates. This would mean that space fluctuates
locally between expansion and contraction with different phases in different
points.

\section{The stress-energy of the vacuum in quantum electrodynamics}

In this section I shall prove that hypothesis $H2$ is indeed falsified by a
straightforward quantum calculation of the stress tensor of the vacuum. I
shall do that for the fields of quantum electrodynamics (QED) but it may be
extended to other quantum fields. Indeed for free Bose fields both the
vacuum density and the pressure are positive, but for free Fermi fields the
density is negative and the pressure positive, as shown in the following. In
the particular case of QED here studied this makes possible that eq.$\left( 
\ref{1}\right) $ holds true, approximately, for the vacuum expectations.

Although it is well known, for later convenience I recall the derivation of
the energy density and pressure of the free electromagnetic field. The
classical Lagrangian of the free field may be written in terms of the
four-vector potential $A_{\mu }$ in the form 
\begin{equation}
\mathcal{L=-}\frac{1}{4}F_{\mu \nu }F^{\mu \nu },F_{\mu \nu }=\frac{\partial
A_{\mu }}{\partial x^{\nu }}.  \label{2.1}
\end{equation}
Hence it is straightforward to get the stress-energy tensor of the free
field. After quantization the operators of the energy density (or
Hamiltonian density), $\hat{\rho}_{EM},$ and the pressure, $\hat{P}_{EM},$
may be written in terms of the elctric, $\mathbf{\hat{E},}$ and the
magnetic, $\mathbf{\hat{B},}$ field operators as follows 
\begin{equation}
\hat{\rho}_{EM}\equiv \frac{1}{2}\left( \mathbf{\hat{E}}^{2}+\mathbf{\hat{B}}%
^{2}\right) =3\hat{P}_{EM}.  \label{2.2}
\end{equation}
Expanding in plane waves we get (ingoring the time dependence) 
\begin{eqnarray}
\hat{\rho}_{EM} &=&\hat{\rho}_{EM0}+\hat{\rho}_{EM1}+\hat{\rho}_{EM2},\hat{%
\rho}_{EM0}=\frac{1}{2V}\sum_{\mathbf{k,\varepsilon }}k,  \nonumber \\
\hat{\rho}_{EM1} &=&\frac{1}{2V}\sum_{\mathbf{k,\varepsilon }}\sum_{\mathbf{k%
}^{\prime }\mathbf{,\varepsilon }^{\prime }}\sqrt{kk^{\prime }}\mathbf{%
\varepsilon }*\mathbf{\varepsilon }^{\prime }\hat{\alpha}_{\mathbf{k}%
^{\prime },\mathbf{\varepsilon }^{\prime }}^{\dagger }\hat{\alpha}_{\mathbf{k%
},\mathbf{\varepsilon }}\exp \left[ i\left( \mathbf{k-k}^{\prime }\right)
\cdot \mathbf{r}\right] ,  \nonumber \\
\hat{\rho}_{EM2} &=&\frac{1}{4V}\sum_{\mathbf{k,\varepsilon }}\sum_{\mathbf{k%
}^{\prime }\mathbf{,\varepsilon }^{\prime }}\sqrt{kk^{\prime }}\mathbf{%
\varepsilon }*\mathbf{\varepsilon }^{\prime }\hat{\alpha}_{\mathbf{k},%
\mathbf{\varepsilon }}\hat{\alpha}_{\mathbf{k}^{\prime },\mathbf{\varepsilon 
}^{\prime }}\exp \left[ i\left( \mathbf{k+k}^{\prime }\right) \cdot \mathbf{r%
}\right] +h.c.,  \label{2.3}
\end{eqnarray}
where $k\equiv \left| \mathbf{k}\right| ,k^{\prime }\equiv \left| \mathbf{k}%
^{\prime }\right| ,\mathbf{\varepsilon }$ is the polarization vector that
fulfils $\mathbf{\varepsilon \cdot k=0}$, and similar for $\mathbf{%
\varepsilon }^{\prime },$ \textit{h.c}. means Hermitean conjugate and for
notational simplicity I have labelled 
\begin{equation}
\mathbf{\varepsilon }*\mathbf{\varepsilon }^{\prime }\equiv \mathbf{%
\varepsilon }\cdot \mathbf{\varepsilon }^{\prime }+\frac{1}{kk^{\prime }}%
\left( \mathbf{k\times \varepsilon }\right) \mathbf{\cdot }\left( \mathbf{k}%
^{\prime }\mathbf{\times \varepsilon }^{\prime }\right) .  \label{2.4}
\end{equation}
In $\hat{\rho}_{EM1}$ I have written the operators in normal order, taking
the commutation relations into account.

The Hamiltonian is obtained by performing a space integral of the energy
density eq.$\left( \ref{2.2}\right) $, that is 
\begin{eqnarray}
\hat{H}_{EM} &=&\lim_{V\rightarrow \infty }\int_{V}\hat{\rho}_{EM}\left( 
\mathbf{r}\right) d^{3}\mathbf{r}=\lim_{V\rightarrow \infty }\int_{V}[\hat{%
\rho}_{EM1}\left( \mathbf{r}\right) +\hat{\rho}_{EM2}\left( \mathbf{r}%
\right) ]d^{3}\mathbf{r}  \nonumber \\
&=&\sum_{\mathbf{k,\varepsilon }}k(\hat{\alpha}_{\mathbf{k},\mathbf{%
\varepsilon }}^{\dagger }\hat{\alpha}_{\mathbf{k},\mathbf{\varepsilon }}+%
\frac{1}{2}).  \label{2.5}
\end{eqnarray}
The term $\hat{\rho}_{EM2}$ does not contribute to the Hamiltonian, and
therefore to the quantum expectation of the vacuum energy, because the space
integral of the latter eq.$\left( \ref{2.2}\right) $ leads to $\mathbf{k}%
^{\prime }\mathbf{=-k}$, whence 
\begin{equation}
\int_{V}\hat{\rho}_{EM2}\left( \mathbf{r}\right) d^{3}\mathbf{r=}\frac{1}{4V}%
\sum_{\mathbf{k,\varepsilon ,\varepsilon }^{\prime }}k\hat{\alpha}_{\mathbf{k%
},\mathbf{\varepsilon }}\hat{\alpha}_{-\mathbf{k},\mathbf{\varepsilon }%
^{\prime }}\mathbf{\varepsilon }*\mathbf{\varepsilon }^{\prime }=0,
\label{2.6}
\end{equation}
as may be easily proved taking eq.$\left( \ref{2.4}\right) $ into account
with the choice $\mathbf{\varepsilon }=\mathbf{\varepsilon }^{\prime }.$

For the free electromagnetic field the vacuum state, $\mid 0\rangle ,$ may
be defined as the state with the minimal energy amongst the eigenvectors of
the operator eq.$\left( \ref{2.5}\right) .$ It is a state with zero photons
and it has the properties 
\begin{equation}
\alpha _{\mathbf{k},\mathbf{\varepsilon }}\mid 0\rangle =0,\langle 0\mid
\alpha _{\mathbf{k},\mathbf{\varepsilon }}^{\dagger }=0,  \label{2.7}
\end{equation}
whence the vacuum expectation of the energy density and the pressure are 
\begin{eqnarray}
\rho _{EM} &=&\lim_{V\rightarrow \infty }\frac{1}{V}\sum_{\mathbf{k}%
}k\rightarrow \left( 2\pi \right) ^{-3}\int kd^{3}k,  \nonumber \\
&=&\frac{1}{2\pi ^{2}}\int_{0}^{k_{\max }}k^{3}dk=\frac{\Lambda ^{4}}{8\pi
^{2}},P_{EM}=\frac{\Lambda ^{4}}{24\pi ^{2}}.  \label{2.8}
\end{eqnarray}
where in the continuous limit an integral has been substituted for the sum
in $\mathbf{k}$, and a cutoff, $\Lambda ,$ has been introduced in the photon
energies.

The stress-energy tensor of the Dirac electron-positron field is\cite
{Schweber} 
\begin{equation}
T^{\mu \nu }=\frac{i}{2}(\bar{\psi}\gamma ^{\nu }\frac{\partial \psi }{%
\partial x_{\mu }}-\frac{\partial \bar{\psi}}{\partial _{\mu }}\gamma ^{\nu
}\psi ),\bar{\psi}\equiv \psi ^{\dagger }\beta ,\gamma ^{0}\equiv \beta
,\gamma ^{k}\equiv \alpha ^{k},  \label{12}
\end{equation}
where $\alpha ^{k}$ and $\beta $ are Dirac's matrices and $\psi $ is the
quantized field. Therefore the energy (or Hamiltonian) density operator $%
\hat{\rho}_{D}$ is 
\begin{equation}
\hat{\rho}_{D}=T_{0}^{0}=\frac{i}{2}\left( \psi ^{\dagger }\frac{\partial
\psi }{\partial x^{0}}-\frac{\partial \psi ^{\dagger }}{\partial x^{0}}\psi
\right) .  \label{13}
\end{equation}
Expansion of the quantum field $\psi $ in plane waves gives, after some
algebra, 
\begin{eqnarray}
\hat{\rho}_{D} &=&\frac{1}{V}\sum_{\mathbf{p},s}\sqrt{m^{2}+\mathbf{p}^{2}}%
\left( b_{\mathbf{p,}s}^{\dagger }b_{\mathbf{p,}s}-d_{\mathbf{p,}s}d_{%
\mathbf{p,}s}^{\dagger }\right) =  \nonumber \\
&=&\frac{1}{V}\sum_{\mathbf{p},s}\sqrt{m^{2}+\mathbf{p}^{2}}\left( b_{%
\mathbf{p,}s}^{\dagger }b_{\mathbf{p,}s}+d_{\mathbf{p,}s}^{\dagger }d_{%
\mathbf{p,}s}-1\right) ,  \label{14}
\end{eqnarray}
where $V$ is the normalization volume, $b_{\mathbf{p,}s}\left( d_{\mathbf{p,}%
s}\right) $ is the electron (positron) destruction operator and $b_{\mathbf{%
p,}s}^{\dagger }(d_{\mathbf{p,}s}^{\dagger })$ the creation operator, $m$
being the electron (or positron) mass, $\mathbf{p}$ its momentum and $s$ its
spin. It is remarkable the quantity $-1$ in the latter form of eq.$\left( 
\ref{14}\right) ,$which is a consequence of the anticommutation rules of
Fermi field operators and gives rise to a negative contribution to the
vacuum energy. The vacuum state, $\mid 0\rangle ,$ consists of zero
electrons and zero positrons. The expectation value of the density operators
eq.$\left( \ref{14}\right) $ gives, after performing the (trivial) sum of
polarizations and passing to the continuous limit $V\rightarrow \infty $,
\begin{eqnarray}
\rho _{D} &=&\langle 0\left| \hat{\rho}_{D}\right| 0\rangle \rightarrow -\pi
^{-2}\int_{0}^{p_{\max }}\sqrt{m^{2}+p^{2}}p^{2}dp=-\pi
^{-2}\int_{m}^{\Lambda }\sqrt{E^{2}-m^{2}}E^{2}dE  \nonumber \\
&=&-\frac{1}{4\pi ^{2}}\left[ \Lambda (\Lambda ^{2}-\frac{1}{2}m^{2})\sqrt{%
\Lambda ^{2}-m^{2}}-\frac{1}{2}m^{4}\cosh ^{-1}\left( \frac{\Lambda }{m}%
\right) \right]   \nonumber \\
&=&-\frac{1}{4\pi ^{2}}\left[ \Lambda ^{4}-\Lambda ^{2}m^{2}-\frac{1}{2}%
m^{4}\ln \left( \frac{\Lambda }{m}\right) \right] +O\left( \frac{m^{6}}{%
\Lambda ^{2}}\right) ,  \label{15}
\end{eqnarray}
where $E=\sqrt{m^{2}+p^{2}}$ and $\Lambda >>m$ is a cutoff energy for the
electron and positron.

In order to get the contribution of the free Dirac field to the vacuum
pressure it is convenient to calculate the expectation value of the trace of
the tensor $T^{\mu \nu }$ eq.$\left( \ref{12}\right) .$ For the quantized
tensor we get 
\begin{eqnarray}
T_{\nu }^{\nu } &=&\frac{i}{2}(\bar{\psi}\gamma ^{\nu }\frac{\partial \psi }{%
\partial x^{\nu }}-\frac{\partial \bar{\psi}}{\partial x^{\nu }}\gamma ^{\nu
}\psi )=\psi ^{\dagger }\left( \hat{H}-\mathbf{\alpha }\cdot \mathbf{\hat{p}}%
\right) \psi =\psi ^{\dagger }m\beta \psi   \nonumber \\
&=&m\bar{\psi}\psi =\frac{1}{V}\sum_{\mathbf{p},s}\frac{m^{2}}{E}\left( b_{%
\mathbf{p,}s}^{\dagger }b_{\mathbf{p,}s}+d_{\mathbf{p,}s}^{\dagger }d_{%
\mathbf{p,}s}-1\right) ,  \label{16}
\end{eqnarray}
where we introduce the differential operators $\hat{H}\equiv i\partial
/\partial t,\mathbf{\hat{p}\equiv -}i\nabla ,$ the third equality takes
Dirac's equation into account and the latter follows after a plane waves
expansion. The vacuum expectation value is 
\begin{eqnarray}
\langle 0\left| T_{\nu }^{\nu }\right| 0\rangle  &=&\langle 0\left| \hat{\rho%
}_{D}+3\hat{P}_{D}\right| 0\rangle =\rho _{D}+3P_{D}=m\langle 0\left| \bar{%
\psi}\psi \right| 0\rangle =-\frac{1}{V}\sum_{\mathbf{p},s}\frac{m^{2}}{E} 
\nonumber \\
&\rightarrow &\frac{m^{2}}{2\pi ^{2}}\left[ -\Lambda ^{2}+m^{2}+m^{2}\ln
\left( \frac{\Lambda }{m}\right) \right] +O\left( \frac{m^{6}}{\Lambda ^{2}}%
\right) ,  \label{17}
\end{eqnarray}
whence 
\begin{equation}
P_{D}=-\frac{1}{3}\rho _{D}+\frac{m^{2}}{6\pi ^{2}}\left[ -\Lambda
^{2}+m^{2}+m^{2}\ln \left( \frac{\Lambda }{m}\right) \right] +O\left( \frac{%
m^{6}}{\Lambda ^{2}}\right) .  \label{18}
\end{equation}
It is interesting that for the electron-positron field the vacuum energy is
negative but the vacuum pressure is positive and from the calculation it may
be realized that this is a general fact for spin-1/2 (Dirac) fields. In
contrast for the electromagnetic field both energy and pressure are positive
and we may conjecture that this is a general fact for Bose fields, although
I omit the proof. The difference derives from the commutation rules (for
Bose fields) vs. anticommutation (for Fermi fields) .

A relevant result is that the stress-energy tensor of the QED vacuum fulfils
eq.$\left( \ref{8}\right) $ modulo terms logaritmically divergent with the
cutoff, therefore of order $m^{4}$ if we assume that a cutoff $\Lambda $
exists, even if it is of order the Planck energy. Indeed we have 
\begin{equation}
\rho _{QED}+P_{QED}=\rho _{EM}+P_{EM}+\rho _{D}+P_{D}=\frac{m^{4}}{12\pi ^{2}%
}\left[ 2+3\ln \left( \frac{\Lambda }{m}\right) \right] .  \label{19}
\end{equation}
This is a remarkable cancelation taking into account the huge values of the
density and the pressure, that are of order $\Lambda ^{4},$ about $%
10^{90}m^{4}$ if $\Lambda $ is the Planck mass. Nevertheless the sum eq.$%
\left( \ref{19}\right) $ is still high, about $10^{35}$ times the dark
energy density.

A quantity of interest is 
\begin{eqnarray}
\rho _{QED}+3P_{QED} &=&\rho _{EM}+3P_{EM}+\rho _{D}+3P_{D}  \nonumber \\
&=&\frac{\Lambda ^{4}}{4\pi ^{2}}-\frac{\Lambda ^{2}m^{2}}{2\pi ^{2}}+\frac{%
m^{4}}{2\pi ^{2}}\left[ \ln \left( \frac{\Lambda }{m}\right) +1\right] ,
\label{19a}
\end{eqnarray}
where eqs.$\left( \ref{2.8}\right) $ and $\left( \ref{18}\right) $ have been
taken into account.

I point out that the total energy density operator of QED should contain
also a term for the interaction, that may be written 
\begin{equation}
\hat{\rho}_{int}\left( \mathbf{r,}t\right) =-e\hat{\psi}^{\dagger }\mathbf{%
\alpha }\hat{\psi}\cdot \mathbf{\hat{A}.}  \label{20}
\end{equation}
The operators $\hat{\psi},\hat{\psi}^{\dagger }$\ and $\mathbf{\hat{A}}$\
contain two terms each when expanded in plane waves, every term
corresponding to an infinite sum. One of these terms has creation operators
and the other one annihilation operators. This gives rise to 8 terms for $%
\hat{\rho}_{int},$\ eq.$\left( \ref{20}\right) $. I will write only the two
terms that will survive in the Hamiltonian. We get (ignoring the time
dependence) 
\begin{eqnarray}
\hat{\rho}_{int}\left( \mathbf{r}\right) &=&\sum_{\mathbf{p,q},\mathbf{k}%
,s,s^{\prime },\varepsilon }\left[ \zeta _{n}\hat{\alpha}_{\mathbf{k},%
\mathbf{\varepsilon }}\hat{b}_{\mathbf{p}s}\hat{d}_{\mathbf{q}s^{\prime
}}\exp \left[ i\left( \mathbf{p+q+k}\right) \cdot \mathbf{r}\right]
+h.c\right]  \nonumber \\
\zeta _{n} &\equiv &-e\frac{m}{V^{1/2}\sqrt{2kEE^{\prime }}}u_{s}^{\dagger
}\left( \mathbf{p}\right) \mathbf{\alpha \cdot \varepsilon }v_{s^{\prime
}}\left( \mathbf{q}\right) ,  \label{21}
\end{eqnarray}
where $h.c.$\ means Hermitean conjugate, $u_{s}^{\dagger }$\textbf{\ }and $%
v_{s^{\prime }}$ are spinors, and $n$ stands for $\left\{ \mathbf{p,q},%
\mathbf{k},s,s^{\prime },\varepsilon \right\} .$ The interaction Hamiltonian 
$\hat{H}_{int}$ is the space integral of $\hat{\rho}_{int}\left( \mathbf{r}%
\right) $ within the volume $V$. One of the terms of the Hamiltonian may
create triples electron-positron-photon and the other term may annihilate
triples. In this paper I will neglect the interaction energy and pressure.
They are typically smaller than the other terms considered in this section
by about $\alpha =1/137.$

\section{Fluctuations in the QED vacuum}

The study of fluctuations requires calculating the stress-energy tensor over
finite regions. For instance we may calculate the vacuum energy density in a
region of volume $v$ using an energy density operator of the form 
\[
\hat{\rho}^{v}\equiv v^{-1}\int_{v}d^{3}\mathbf{r}\hat{\rho}\left( \mathbf{r}%
\right) \rightarrow \rho ^{v}=\langle vac\mid \hat{\rho}^{v}\mid vac\rangle
, 
\]
where $\hat{\rho}\left( \mathbf{r}\right) $ is the quantum operator for the
energy density (see e. g. eq.$\left( \ref{2.3}\right) )$. At a difference
with section 2 where a similar calculation involved the limit $V\rightarrow
\infty $, see eq.$\left( \ref{2.8}\right) ,$ here the volume $v$ remains
finite. The consequence is that the density fluctuates in the sense that 
\[
\langle vac\mid \left( v^{-1}\int_{v}d^{3}\mathbf{r}\hat{\rho}\right)
^{2}\mid vac\rangle >\langle vac\mid v^{-1}\int_{v}d^{3}\mathbf{r}\hat{\rho}%
\mid vac\rangle ^{2}, 
\]
that would become an equality in the limit $v\rightarrow \infty .$ Similar
results hold for the pressure.

The most interesting quantity for us is the sum $\rho _{QED}+3P_{QED}$ of
components of the vacuum stress-energy tensor, see eq.$\left( \ref{2.11}%
\right) $. In order to study the fluctuations we define the folowing
operator 
\begin{equation}
\hat{Q}\equiv v^{-1}\int_{v}d^{3}\mathbf{r[}\hat{\rho}_{EM}+3\hat{P}_{EM}+%
\hat{\rho}_{D}+3\hat{P}_{D}],  \label{3.2}
\end{equation}
where eqs.$\left( \ref{2.2}\right) ,\left( \ref{2.3}\right) $ and $\left( 
\ref{17}\right) $ have been taken into account. The probability distribution
predicted by quantum mechanics for the quantity Q, associated to the
operator $\hat{Q}$, is 
\begin{equation}
F(Q)=\frac{1}{2\pi }\int d\lambda \exp \left( -i\lambda Q\right) \langle
vac\mid \exp \left( i\lambda \hat{Q}\right) \mid vac\rangle ,  \label{3.1}
\end{equation}
It is easy to see that $\hat{\rho}_{EM}+3\hat{P}_{EM}$ commutes with $\hat{%
\rho}_{D}+3\hat{P}_{D}$, the former involving electromagnetic field
operators and the latter electron and positron operators. Furthermore it may
be realized that the expectation eq.$\left( \ref{17}\right) $ is actually an
eigenvalue of the operator $\hat{\rho}_{D}+3\hat{P}_{D}$ with eigenvector $%
\mid 0\rangle .$ Therefore we may rewrite eq.$\left( \ref{3.1}\right) $ in
the form 
\begin{eqnarray}
F(Q) &=&\frac{1}{2\pi }\int d\lambda \exp \left\{ i\lambda \left[
v^{-1}\int_{R}d^{3}\mathbf{r}\left( \rho _{D}+3P_{D}\right) -Q\right]
\right\} S  \nonumber \\
S &\equiv &\langle 0\mid \exp \left\{ i\lambda \left[ v^{-1}\int_{R}d^{3}%
\mathbf{r}\left( \hat{\rho}_{EM}+3\hat{P}_{EM}\right) \right] \right\} \mid
0\rangle ,  \label{3.3}
\end{eqnarray}
where I approximate $\mid vac\rangle $ by $\mid 0\rangle .$ The vacuum
ground state $\mid vac\rangle ,$ including the interaction eq.$\left( \ref
{21}\right) $, is different from the ground state $\mid 0\rangle $ of the
free fields alone, but our approximation in eq.$\left( \ref{3.3}\right) $ is
consistent with neglecting the interaction, as was made in section 2. Taking
into account eq.$\left( \ref{2.3}\right) $ we may write 
\begin{eqnarray}
\hat{\rho}_{EM}+3\hat{P}_{EM} &=&2\hat{\rho}_{EM}=\frac{1}{2V}\sum_{\mathbf{%
k,\varepsilon }}\sum_{\mathbf{k}^{\prime }\mathbf{,\varepsilon }^{\prime }}%
\sqrt{kk^{\prime }}\mathbf{\varepsilon }*\mathbf{\varepsilon }^{\prime } 
\nonumber \\
&&\times \left[ \hat{\alpha}_{\mathbf{k},\mathbf{\varepsilon }}\exp \left( i%
\mathbf{k}\cdot \mathbf{r}\right) \mathbf{+}\hat{\alpha}_{\mathbf{k},\mathbf{%
\varepsilon }}^{\dagger }\exp \left( -i\mathbf{k}\cdot \mathbf{r}\right)
\right]  \nonumber \\
&&\times \left[ \hat{\alpha}_{\mathbf{k}^{\prime },\mathbf{\varepsilon }%
^{\prime }}\exp \left( i\mathbf{k}^{\prime }\cdot \mathbf{r}\right) +\hat{%
\alpha}_{\mathbf{k}^{\prime },\mathbf{\varepsilon }^{\prime }}^{\dagger
}\exp \left( -i\mathbf{k}^{\prime }\cdot \mathbf{r}\right) \right] .
\label{3.4}
\end{eqnarray}

Calculating accurately the probability density $F(Q)$ via eq.$\left( \ref
{3.3}\right) $ would be cumbersome. In fact\textrm{\ }we should expand $S$
in powers of $\lambda ,$ that is redefine 
\begin{equation}
S\equiv 1+\sum_{n=1}^{\infty }\frac{(i\lambda )^{n}}{n!}\langle 0\mid \left(
v^{-1}\int_{R}2\hat{\rho}_{EM}d^{3}\mathbf{r}\right) ^{n}\mid 0\rangle ,
\label{3.5}
\end{equation}
then calculate the general term and finally sum the series. In addition the
interaction energy operator, $\hat{\rho}_{int},$ should be included in the
vacuum expectation. In this paper I will not make an accurate calculation
but an estimate of $F(Q)$ using some approximations as explained in the
following.

I will assume that the probability density $F(Q)$ is a Gaussian that is 
\begin{equation}
F(Q)=\frac{1}{\sqrt{2\pi }\sigma }\exp \left[ -\frac{\left( Q-Q_{0}\right)
^{2}}{2\sigma ^{2}}\right] ,  \label{3.10}
\end{equation}
and get the parameters $Q_{0}$ and $\sigma $ via identifying 
\[
\left\langle Q\right\rangle \equiv \int QF(Q)dQ=\langle 0\mid \hat{Q}\mid
0\rangle ,\left\langle Q^{2}\right\rangle =\langle 0\mid \hat{Q}^{2}\mid
0\rangle , 
\]
whence taking eq.$\left( \ref{19a}\right) $ into account we get 
\begin{eqnarray}
Q_{0} &=&\langle 0\mid v^{-1}\int_{R}d^{3}\mathbf{r[}2\hat{\rho}_{EM}+\hat{%
\rho}_{D}+3\hat{P}_{D}]\mid 0\rangle =\rho _{D}+3P_{D}+2\rho _{EM}, 
\nonumber \\
&=&\frac{1}{4\pi ^{2}}\left[ \Lambda ^{4}-2\Lambda ^{2}m^{2}+2m^{4}\left[
\ln \left( \frac{\Lambda }{m}\right) +1\right] \right] .  \label{3.6}
\end{eqnarray}
The former equality may be checked easily from the expressions of the
density and pressure operators, eqs.$\left( \ref{12}\right) $ to $\left( \ref
{14}\right) $.

In order to get the variance of the distribution, $\sigma ^{2},$ it is
convenient to use eq.$\left( \ref{2.3}\right) $, that is 
\begin{eqnarray}
\sigma ^{2} &=&v^{-2}\langle 0\mid (\int_{R}2\hat{\rho}_{EM}d^{3}\mathbf{r}%
)^{2}\mid 0\rangle -\left( 2\rho _{EM}\right) ^{2}  \nonumber \\
&=&v^{-2}\langle 0\mid (\int_{R}2\hat{\rho}_{EM2}d^{3}\mathbf{r})^{2}\mid
0\rangle =\frac{v^{-2}}{8V^{2}}\sum_{\mathbf{k,\varepsilon }}\sum_{\mathbf{k}%
^{\prime }\mathbf{,\varepsilon }^{\prime }}kk^{\prime }(\mathbf{\varepsilon }%
*\mathbf{\varepsilon }^{\prime })^{2}  \nonumber \\
&&\times (\hat{\alpha}_{\mathbf{k},\mathbf{\varepsilon }}^{\dagger }\hat{%
\alpha}_{\mathbf{k}^{\prime },\mathbf{\varepsilon }^{\prime }}^{\dagger }%
\hat{\alpha}_{\mathbf{k},\mathbf{\varepsilon }}\hat{\alpha}_{\mathbf{k}%
^{\prime },\mathbf{\varepsilon }^{\prime }})\int_{v}d^{3}\mathbf{r}%
\int_{v}d^{3}\mathbf{r}^{\prime }\exp \left[ i\left( \mathbf{k+k}^{\prime
}\right) \cdot (\mathbf{r-r}^{\prime })\right]  \nonumber \\
&=&\frac{v^{-2}}{4V^{2}}\sum_{\mathbf{k}}\sum_{\mathbf{k}^{\prime
}}kk^{\prime }\left( 1+\frac{\mathbf{k\cdot k}^{\prime }}{kk^{\prime }}%
\right) ^{2}\left| \int_{v}d^{3}\mathbf{r}\exp \left[ i\left( \mathbf{k+k}%
^{\prime }\right) \cdot \mathbf{r}\right] \right| ^{2},  \label{3.7}
\end{eqnarray}
where a sum of polarizations has been performed in the latter equality. I
point out that $v$ is a small volume, say with size the Planck length,
whilst $V$ is an integration volume introduced for calculational
convenience. In fact it allows treating the fields via an expansion in plane
waves, putting $V\rightarrow \infty $ at some stage of the calculation in
order to substitute integrals for the more involved sums.

For the (small) region with volume $v$ I choose a sphere of radius $R$ so
that 
\begin{equation}
v=\frac{4}{3}\pi R^{3}.  \label{v}
\end{equation}
Then the space integral in eq.$\left( \ref{3.7}\right) $ is straightforward
and we get 
\begin{equation}
I\equiv \left| \int_{R}d^{3}\mathbf{r}\exp \left[ i\left( \mathbf{k+k}%
^{\prime }\right) \cdot \mathbf{r}\right] \right| ^{2}=4\pi ^{2}R^{6}\frac{%
\left( \sin x-x\cos x\right) ^{2}}{x^{6}},  \label{3.8}
\end{equation}
where 
\begin{equation}
x\equiv R\left| \mathbf{k+k}^{\prime }\right| =R\sqrt{k^{2}+k^{\prime
2}+2kk^{\prime }u},u\equiv \cos \theta ,  \label{3.9}
\end{equation}
$\theta $ being the angle between $\mathbf{k}$ and $\mathbf{k}^{\prime }.$
Thus eq.$\left( \ref{3.7}\right) $ becomes, in the continuous limit $%
V\rightarrow \infty $, 
\begin{equation}
\sigma ^{2}=\frac{1}{\left( 16\pi ^{3}\right) ^{2}}\int_{0}^{\Lambda }4\pi
k^{3}dk\int_{0}^{\Lambda }2\pi k^{\prime 3}dk^{\prime }\int_{-1}^{1}du\left(
1+u\right) ^{2}I\left( x\right) ,  \label{3.11}
\end{equation}
where $\Lambda $ is the cutoff defined in section 2. It is convenient to go
from the variable $u$ to $x$ whence we get 
\begin{eqnarray*}
\sigma ^{2} &=&\frac{8v^{-2}}{\left( 16\pi ^{2}\right) ^{2}}%
\int_{0}^{\Lambda }k^{3}dk\int_{0}^{\Lambda }k^{\prime 3}dk^{\prime
}\int_{R\left| k-k^{\prime }\right| }^{R\left| k+k^{\prime }\right| }\frac{%
xdx}{R^{2}kk^{\prime }}I\left( x\right) \\
&&\times \left[ 1+\frac{x^{2}-R^{2}\left( k^{2}+k^{\prime 2}\right) }{%
2kk^{\prime }R^{2}}\right] ^{2},
\end{eqnarray*}
that is, with the new change of variable $y=x^{2}$, 
\begin{equation}
\sigma ^{2}=\frac{v^{-2}}{\left( 4\pi \right) ^{4}R^{6}}\int_{0}^{\Lambda
}dk\int_{0}^{\Lambda }dk^{\prime }\int_{R^{2}\left( k-k^{\prime }\right)
^{2}}^{R^{2}\left( k+k^{\prime }\right) ^{2}}dyI\left( \sqrt{y}\right)
\left[ y-R^{2}\left( k-k^{\prime }\right) ^{2}\right] ^{2}.  \label{3.12}
\end{equation}

The integral is involved and I will simplify it approximating $I\simeq (4\pi
^{2}R^{6}/9)/(1+2y^{2}),$ that agrees with eq.$\left( \ref{3.8}\right) $ in
both limits $y<<1$ and $y>>1$. Thus we get 
\begin{equation}
\sigma ^{2}=\frac{R^{2}v^{-2}}{9\times 32\pi ^{2}}\int_{0}^{\Lambda
}kdk\int_{0}^{\Lambda }k^{\prime }dk^{\prime }J(k,k^{\prime }),  \label{3.14}
\end{equation}
where 
\begin{eqnarray*}
J &\equiv &\left[ \frac{y}{2}+\frac{R^{2}\left( \left( k-k^{\prime }\right)
^{2}-1/2\right) }{\sqrt{2}}\arctan \left( \sqrt{2}y\right) -R^{2}\left(
k-k^{\prime }\right) ^{2}\ln \left( 1+2y^{2}\right) \right] _{R^{2}\left(
k-k^{\prime }\right) ^{2}}^{R^{2}\left( k+k^{\prime }\right) ^{2}} \\
&=&2R^{2}kk^{\prime }-R^{2}\left( k-k^{\prime }\right) ^{2}\ln \frac{%
1+2R^{4}\left( k+k^{\prime }\right) ^{4}}{1+2R^{4}\left( k-k^{\prime
}\right) ^{4}} \\
&&+\frac{R^{2}\left( \left( k-k^{\prime }\right) ^{2}-1/2\right) }{\sqrt{2}}%
\left[ \arctan \left( \sqrt{2}R^{2}\left( k+k^{\prime }\right) ^{2}\right)
-\arctan \left( \sqrt{2}R^{2}\left( k-k^{\prime }\right) ^{2}\right) \right]
.
\end{eqnarray*}
The integrals are lenghty and finding an accurate result is not worth taking
into account the other approximations involved. I will report only an upper
bound obtained retaining the positive terms of $J$ alone. This provides a
rough approximation taking into account that  $J>0$ as it follows from eq.$%
\left( \ref{3.7}\right) .$ Thus from the inequality 
\[
\pi /2>\arctan \left( \sqrt{2}R^{2}\left( k+k^{\prime }\right) ^{2}\right)
-\arctan \left( \sqrt{2}R^{2}\left( k-k^{\prime }\right) ^{2}\right) >0,
\]
we estimate 
\begin{equation}
\sigma ^{2}\lesssim \frac{R^{4}v^{-2}}{9\times 32\pi ^{2}}\int_{0}^{\Lambda
}kdk\int_{0}^{\Lambda }k^{\prime }dk^{\prime }\left[ 2kk^{\prime }+\frac{\pi 
}{2\sqrt{2}}\left( k-k^{\prime }\right) ^{2}\right] \simeq \frac{\Lambda
^{6}R^{4}}{3200v^{2}}.  \label{3.15}
\end{equation}
Taking eq.$\left( \ref{v}\right) $ into account we get finally

\begin{equation}
\sigma ^{2}\lesssim \frac{\Lambda ^{6}R^{4}}{3200}\left( \frac{3}{4\pi R^{3}}%
\right) ^{2}.  \label{3.16}
\end{equation}

\section{The accelerating expansion of the universe}

From eq.$\left( \ref{2.11}\right) $ we see that when $\rho +3P>0$ the space
oscillates rapidly between expansion and contraction, but when $\rho +3P<0$
it expand exponentially, that is 
\begin{equation}
a=a_{0}\exp \left( Ht\right) ,H=\sqrt{\frac{4\pi }{3}G\left( \rho +3P\right) 
}\text{ if }\rho +3P<0.  \label{3.13}
\end{equation}
As is well known no satisfactory quantum gravity theory is available and the
popular method to treat problems involving simultaneously quantum theory and
gravity is the semiclassical approximation. It consists of using as
stress-energy tensor, in the right side of Einstein\'{}s equation, the
expectation of the quantized tensor in the appropriate quantum state, here
the vacuum state $\mid 0\rangle $. In order to take account of the quantum
fluctuations it is necessary to extend the semiclassical approximation as
follows. I propose to treat $F(Q),$ eq.$\left( \ref{3.1}\right) $ as a
probability distribution of the trace, $\rho _{QED}+3P_{QED},$ of the
stress-energy tensor. This amounts at applying the semiclassical
approximation to all expectations $\langle 0\mid \left( v^{-1}\int_{R}2\hat{%
\rho}_{EM}d^{3}\mathbf{r}\right) ^{n}\mid 0\rangle ,$ eq.$\left( \ref{3.5}%
\right) .$ After that Einstein equation provides, in principle, a
probability distribution of metrics. Of course making a calculation of this
probability distribution would be a formidable task, but here our more
modest task will be to estimate the probability that the metric is of the
form of eq.$\left( \ref{3.13}\right) $ in a small region where we assume
that stress-energy tensor is homogeneous and isotropic. In fact I propose to
estimate the Hubble parameter $H$ by calculating the expectation (in the
sense of classical probability theory) that $Q=\rho +3P<0$ from the
probability distribution eq.$\left( \ref{3.10}\right) $. That is I assume
that the effective $Q$ for the accelerating expansion is given by the
integral 
\begin{eqnarray}
Q_{eff} &=&\int_{-\infty }^{0}QF(Q)dQ=\frac{\sigma }{\sqrt{2\pi }}%
\int_{-\infty }^{-\frac{Q_{0}}{\sigma }}\left( x+\frac{Q_{0}}{\sigma }%
\right) \exp \left( -\frac{x^{2}}{2}\right) dx  \label{4.14} \\
&=&\frac{\sigma }{\sqrt{2\pi }}\left[ -\exp \left( -\frac{Q_{0}^{2}}{2\sigma
^{2}}\right) +\frac{Q_{0}}{\sigma }erfc\left( -\frac{Q_{0}}{\sqrt{2}\sigma }%
\right) \right]   \nonumber \\
&\simeq &\frac{\sigma }{\sqrt{2\pi }}\times \frac{Q_{0}}{\sigma }\times \exp
\left( -\frac{Q_{0}^{2}}{2\sigma ^{2}}\right) \simeq \frac{Q_{0}}{\sqrt{2\pi 
}}\exp \left( -\frac{Q_{0}^{2}}{2\sigma ^{2}}\right) .  \nonumber
\end{eqnarray}
Approximating $Q_{0}$ by the leading term in eq.$\left( \ref{3.6}\right) ,$
taking eq.$\left( \ref{3.15}\right) $ into account,we get 
\[
Q_{0}\simeq \frac{\Lambda ^{4}}{4\pi ^{2}},\frac{Q_{0}^{2}}{2\sigma ^{2}}>%
\frac{3200\Lambda ^{2}R^{2}}{2\times \left( 3\pi \right) ^{2}}=18\Lambda
^{2}R^{2},
\]
whence 
\[
Q_{eff}<\frac{\Lambda ^{4}}{\left( 2\pi \right) ^{5/2}}\exp \left(
-18\Lambda ^{2}R^{2}\right) .
\]
This would fit in the observed value of the dark energy needed to explain
the accelerated expansion of the universe if we choose $R\sim 4\Lambda ^{-1}.
$ In fact in this case taking eq.$\left( \ref{4.14}\right) $ into account we
would get 
\[
\frac{Q_{0}^{2}}{2\sigma ^{2}}=288\Rightarrow Q_{eff}\lesssim Q_{0}\exp
\left( -284\right) =10^{-123}\Lambda ^{4}.
\]
That is, assuming that $\Lambda ^{4}$ is the Planck density, then $R$ needs
to be about $4$ times the Planck length.

\section{Conclusions}

I argue that a solution to the cosmological constant problem is to assume
that the expectation value of the quantum vacuum stress-energy tensor is
proportional to the metric tensor with a negative energy density and
positive pressure. Indeed this fits with the results of a straightforward
calculation of the said expectation for free Bose and Fermi fields, at least
for the fields of quantum electrodynamics (electromagnetism and
electron-positron Dirac field). As a consequence the metric of the universe
might corresponds to a FLRW with accelerated expansion only after averaging
on large scales. However at small scales it gives rise to an extremely rapid
fluctuation between expansion and contraction in every small region, with
different phases in different points. Of course both our assumption and the
result of the calculation in QED contradict the standard wisdom that the
vacuum energy density should be positive. This puts a difficulty that should
be further studied.

The vacuum stress-energy tensor has huge fluctuations of the same order of
its average value. This fact implies that some specially big fluctuations
may give rise to a tensor with positive energy and negative pressure,
leading to short periods of expansion in every small region. The very small
probability of such extremely large fluctuations implies that the overall
expansion of the universe is very slow. An explicit calculation with
plausible approximations leads to an estimate that fits in the observed
value of the accelerated expansion.

\textit{Acknowledgement}. I thank Luis de la Pe\~{n}a for pointing out to me
the relevance of the paper by Wang et al.\cite{Wang}.

\end{document}